# Symmetry and dissipation as the basic mechanism of social mobility, explaining distance scaling of migration patterns

Kirill S. Glavatskiy

*Centre for Complex Systems, The University of Sydney, Australia.*

Models of social mobility inspired by the Newton's law of gravity have been used for several decades to describe migrations of people, goods, and information. Despite an eminent reference and widespread use, these models lack the background theory, being often viewed as a collection of empirical recipes which rely on adjustable parameters. Here we propose a tractable and fundamental mechanism of social mobility, which explains distance scaling of migration flows and predicts the value of the scaling exponent. The mechanism reveals two key aspects framing social flows, which have direct analogy in physics: symmetry and dissipation. In particular, we identify the conditions for the social "gravity" scaling, when the power law exponent equals 2, and explain deviations from this behaviour, including saturation transitions. The resulting flow distribution is determined by the spatial structure of the underlying social network, rather than by distance explicitly. The theory is verified for residential migration in suburb networks of major Australian cities with diverse structure, population and size. The mechanism is directly translatable to other social contexts, such as flows of goods or information, and provides a universal understanding of how dynamic patterns of social mobility emerge from structural network properties.

*e-mail: k.s.glavatskiy@gmail.com

Newton's law of gravity offers a rigorous and universal expression for the distance dependence of the force between two massive bodies, irrespectively of the circumstances they interact in. A similar distance-dependent expression has been proposed in social sciences [1, 2], describing the migration of people between certain locations [3–8], as well as flow of goods [9–13] or information [14–16]. In particular, according to various social gravity models, the flow $F_{ij}$ between two sites $i$ and $j$ which are separated by the Euclidean distance $r_{ij}$ and have the populations $P_i$ and $P_j$ respectively is expressed as

$$F_{ij} = k \frac{P_i P_j}{f(r_{ij}; \gamma)} \tag{1}$$

where $f(r_{ij}; \gamma)$ is a deterrence function. In the physical gravity law $f = r_{ij}^\gamma$ with $\gamma = 2$, while in social gravity models $\gamma$ is an adjustable parameter chosen to fit empirical data. Empirical deterrence function may also have a non-power law form, e.g. $f = \exp(\gamma r_{ij})$.

Despite the widespread practical use [17, 18], the fundamental nature of the social gravity law is often questioned [19–22]. First, unlike the physical gravity law, the social gravity law (1) lacks a rigorous derivation, being merely an empirical observation. It is typically referred to by the term "gravity models", emphasizing existence of a range of specific models which work in specific circumstances. In particular, the social gravity law is believed to have limited range of theoretical applicability, and alternative models have been proposed to address those limitations [19, 22]. Second, variation of the parameters used in these models across specific applications limits the universality of social gravity law. In particular, while in the physical gravity law the deterrence function is exactly $1/r^2$, in social gravity laws the values of the exponent range from 0.3 to 3.0. Third, in its power-law form the social migration flux increases to infinity at small distances, while in reality migration plateaus, reaching a certain saturation level. Community clustering and the corresponding network spatial structure has been shown to alter this simple scaling behaviour [23–25]. Finally, the social gravity law is formulated in a deterministic and not statistical terms. This implies existence of strict mobility rules in social systems, which are hard to expect from largely stochastic systems [26, 27].

In this paper we propose the mechanism, which explains the origin of the social gravity law and identifies the conditions of its applicability. This mechanism does not suffer from empiricism and exploits the fundamental features of spatial organization of the underlying social network in a two-dimensional space. The network sites exchange agents (individual persons, goods, information pieces), which results in their collective flow on the network. We show that the resulting flow distribution between the sites is determined by the spatial and topological structure of the network. We illustrate the predictions of the mechanism for residential migration in eight Australian Greater Capital areas, urban characteristics of which span a broad range of values, with the underlying network represented by residential suburbs.

**Conditions**

Consider a system of agents, which are able to move between the sites of a network. Each agent possesses a certain "energy" which is expended during relocation. Their collective behaviour is formulated in mean-field terms, assuming the agents are indistinguishable and follow a certain statistical distribution. Let $Q_i^+$ and $Q_i^-$ denote the number of agents moving from and to site $i$, which define the source and the sink flow, respectively. Furthermore, let $F_{ij}$ denote the number of agents moving from site $i$ to site $j$, defining the flow between them. Then, the average flow between two sites separated by the distance $r$ is described as

$$\langle \frac{F_{ij}}{Q_i^+ Q_i^-} \rangle = \frac{1}{Q} \frac{\eta}{\nu + N(r)} \tag{2}$$

Here $Q$ is the network-average sink/source flow, $N(r)$ is the average number of sites inside the circle of radius $r$, while $\eta$ and $\nu$ are topological parameters of the network defined below.

The social gravitation mechanism, which results in Equation (2), is illustrated in Figure 1 and described below. Figure 2 shows the predictions of Equation (2) along with the actual data. Notably, not only Equation (2) reveals the "power law" behaviour at moderate distances, but it also reproduces "saturation" transitions at low and high distances. The social gravitation mechanism requires the following conditions to be satisfied.

*Condition I (Symmetry):*

*(a)* The network sites are indistinguishable, except for their position in the two-dimensional metric space and source/sink capacities. This means, in particular, that sites do not have any explicit "attractiveness" which could drive migration. This also means that the source/sink flows are uncorrelated across different sites.

*(b)* The number of agents is conserved, i.e. the total source flow from all source sites of the network is equal to the total sink flow to all destination sites of the network.

*Condition II (Dissipation):*

*(a)* The agents are, in general, bound to a source site, and they need to overcome a certain "potential barrier" to flow outside. Agents with the energy below the potential barrier remain immobile.

*(b)* The excess energy of mobile agents, remained after passing the potential barrier, is "dissipated" in the links connecting the network sites. Each intermediate link between a source and a destination site exerts the same amount of "resistance" on the flow.

Unlike Equation (2), gravity models are typically formulated explicitly in terms of the site populations $P_i$ and a power-law deterrence function. This requires additional conditions to be satisfied, which translate the source and sink flows at each site to its population and quantify the underlying spatial structure of the network, in particular:

*Condition III (Mobility)*:

Both the total source flow and the total sink flow at every site is proportional to the population of that site, with the same proportionality coefficient across all sites, so that $Q_i^+ = \alpha^+ P_i$ and $Q_i^+ = \alpha^- P_i$, where $\alpha^+ = \alpha^- \equiv \alpha$ (because $\sum_i Q_i^+ = \sum_i Q_i^-$) measures the intensity of the population mobility.

*Condition IV (Space)*:

The spatial density of the network sites is a power function of the distance. As the area is proportional to $r^2$, this ensures that average amount of sites in that area is $N(r) \sim r^{2+\delta}$, with $\delta$ indicating how far from homogeneous the spatial distribution of the network sites is. In particular, for a spatially uniform distribution $\delta = 0$, so that $N(r) \sim r^2$.

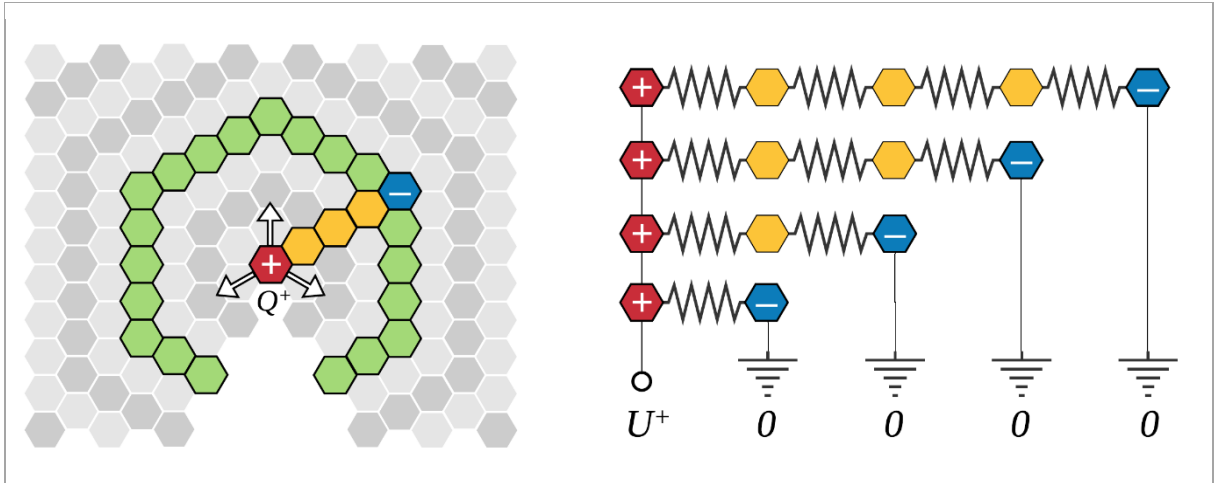

**Figure 1 | Social gravitation mechanism**. Left panel: Schematic representation of a migration network, with the nodes illustrated by a pattern of tiles. The sites are labelled as following: a source ("+", red), a destination ("–", blue), perimetric (green), radial (yellow). The bottom blank sector denotes irregularities of mapping the actual sites with various size and shape to a regular pattern. Right panel: schematic representation of the dissipation mechanism, illustrated by an electric circuit. The source site is shown as the voltage source, the destination sites are shown as ground points, and the links between sites are shown as resistors.

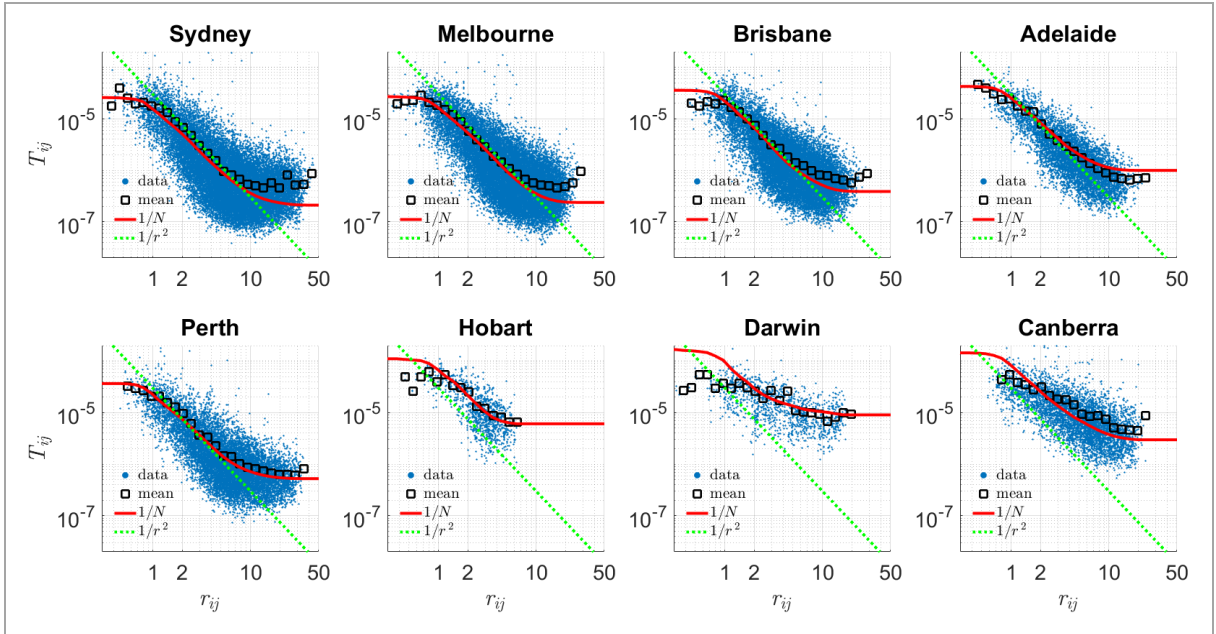

**Figure 2 | Distance dependence of the migration flow**. Each panel shows migration flows in a single Australian Greater Capital area (GCA), according to 2016 Census data of 5-years residential relocation. The properties of each GCA are listed in Table 1. Each blue dot in the cloud represents the actual migration flow quotient $T_{ij} \equiv F_{ij}/(Q_i^+ Q_j^-)$ between each pair of suburbs plotted against the relative distance $r_{ij} \equiv 2 d_{ij}/(\sqrt{S_i} + \sqrt{S_j})$ between these suburbs, where $d_{ij}$ is the Euclidian distance between centroids of the suburbs, while $S_i$ is the area of the suburb. The red line shows the prediction of Equation (2), while the black squares show the average of the actual migration flow over the suburb pairs for which the flow is nonzero. The dotted green line shows the function $a/r^2$ with the same value $a = 3 \cdot 10^{-5}$ for all GCA. The predictions of Equation (2) closely resemble the average value of the actual migration flow for all distances. In contrast, a plain gravity model gives only partial resemblance, diverging from the actual data in the value of the scaling exponent and in the deterrence function for low and high distances.

| 4

The additional conditions lead to an ordinary gravity model for the social migration flow

$$\langle \frac{F_{ij}}{P_i P_j} \rangle = \frac{\kappa}{r^{\gamma}} \tag{3}$$

with $\kappa = \eta \alpha^2 / Q$ and $\gamma = 2 + \delta$. Note, that the form of Equation (3) requires also that the sites' potential barrier is zero.

Figure 3 shows the actual variation of the migration quotient, $F_{ij}/(P_i^+ P_j^-)$, with the distance, as well as the radial distribution functions of the network sites, $N(r)$. It confirms, for residential relocation, that when the additional Conditions III and IV are satisfied, Equation (3) is also valid.

**Mechanism**

We identify the flow variables, $Q^+$, $Q^-$, $E_+$, $F_{+-}$, $J_+$, each denoting the number of agents moving to and from the network sites in a particular combination.

*First*, let $Q^+$ be the total flow from a source site, and $Q^-$ be the total flow to a destination site. Each site can be a source and a destination, resulting in a flow $F_{+-}$ between each pair of sites. We identify the flow field $E_+(-)$ which shows the *potential* flow from a source site, "+", at the location of a destination site, "−", irrespectively of its sink capacity. By definition, the total potential flow at all destinations is equal to the source flow: $\sum_- E_+(-) = Q^+$. Unlike the potential flow $E_+(-)$, the *actual* flow $F_{+-}$ accounts for the destination sink capacity $Q^-$. Because all sites are indistinguishable, the flow from each source site is distributed among all destination sites proportionally to their sink capacity: $F_{+-} = E_+ Q^-/\langle Q^- \rangle$, where $\langle Q^- \rangle = \langle Q^+ \rangle \equiv Q$ is the average sink/source flow in the network.

*Second*, we identify the *energy* $U_+$ of the source flow, which is dissipated in the links connecting the intermediate sites. When the flow reaches a destination site, its energy becomes zero, irrespectively of the distance, similarly to potential of the ground point in electrical circuits. Furthermore, we identify $N_P$ perimetric sites, which are separated from the source site by the same number $N_r$ of radial sites as the destination site. Let $J_+(r)$ be the migration *flux*, which shows the potential flow from a source site to the entire layer of perimetric destination sites. By definition, the sum of the migration fluxes across all perimetric layers is equal to the total source flow: $\sum_r J_+(r) = Q^+$. Because each intermediate link along the radial direction exerts the same amount of resistance on the flow, the flux $J_+(r) = U_+/N_r$. If follows from flow conservation, that $\sum_{N_r}(U_+/N_r) = Q^+$, so the source energy $U_+ = Q^+/H_\theta$, where $H_n \equiv \sum_{k=1}^{n}(1/k)$ is the $n$-th harmonic number and $\theta$ is the half of the network average path length.

*Third*, we observe that the migration flux is the sum of the potential flows at the sites, which constitute a single perimetric layer: $J_+(r) = \sum_p E_+(p)$. Because the perimetric sites are indistinguishable, the flux $J_+ = N_P E_+$, therefore $E_+ = J_+/N_P$. Substituting $J_+$ and $U_+$, we obtain $E_+(r) = Q^+/(H_\theta N_r N_P)$. Note, that the flow field $E_+(r)$ is the same at all perimetric destination sites from a single perimetric layer.

*Fourth*, we estimate $N_P N_r = 2(N - 1)$, where $N$ is the total amount of sites inside the "circle" bounded by the corresponding perimetric sites. This expression is analogous to the geometric relation $p\, r = 2S$ between the circle area $S$, its radius $r$ and perimeter $p$, with the source site being subtracted from the total number of sites, giving the total number of destination sites.

*Fifth*, we identify the "potential barrier" $V$ of a source site, which reduces the energy of the source flow. This barrier essentially increases the effective number of the dissipative links, $N_P N_r$, by the number of neighbouring sites.

*Finally*, collecting the above expressions, we obtain for the potential flow $E_+(r) = H_\theta^{-1} Q^+/[V + 2(N(r) - 1)]$, and for the actual flow $F_{+-}(r) = \eta\, Q^+ Q^-/[(\nu + N(r))\langle Q^- \rangle]$ with $\eta \equiv 1/(2H_\theta)$ and $\nu \equiv V/2 - 1$. Upon averaging the relative flow quotient $F_{+-}/(Q^+ Q^-)$ across all pairs of source and destination sites, we obtain Equation (2).

## $1/N$ scaling and power-2 law

Newton's law of gravity allows a simple interpretation in terms of the gravitational field $E$, defined as the gravitational force exerted on a test unit mass. Namely, according to the Gauss's law for gravity, any mass is the source of the gravitational field, and the total gravitational flux $J$ through any closed surface $\sigma$ is proportional to the mass $m$ enclosed within that surface. By definition, this flux is the surface integral of the field, $J = \int E\, d\sigma$, which in three-dimensional space with spherical symmetry reduces to $J = E\, 4\pi r^2$. This results in the gravitational field $E \sim m/r^2$, so the gravitational force between two bodies with masses $m_i$ and $m_j$ separated by the distance $r_{ij}$ is $F_{ij} \sim m_i m_j / r_{ij}^2$. The Gauss' theorem is a mathematical mechanism which reflects the physical principle of matter conservation. It also finds applications in other areas of physics, resulting in Coulomb's law in electrostatics, the continuity equation in hydrodynamics, and the inverse-square law in radiation.

In analogy, we introduce the *migration field E* and hence formulate *the Gauss' law of social migration*, as following: the total flux $J_i$ of the migration field $E_i$ from the site $i$ to all sites along a closed perimeter $\ell$ around $i$ is proportional to the source flow $Q_i^+$ from that site.

Social migration takes place on a two-dimensional manifold (surface of Earth), hence the migration flux is measured through a one-dimensional manifold (perimeter). This leads to the scaling behaviour for the migration filed $E_i \sim Q_i^+ /r$ and the corresponding migration flow $F_{ij} \sim Q_i^+ Q_j^- / r_{ij}$, which implies the migration exponent $\gamma = 1$. The formulated Gauss' mechanism is universal, resulting in the same value of the migration exponent in all social contexts. However, as it was mentioned earlier, the observed value of this exponent varies across applications and also differs significantly from 1. This means that, unlike for physical gravitation, the Gauss' mechanism alone is not sufficient to describe social migration. The additional factors, which affect the migration scaling behaviour, are migration dissipation and spatial distribution of the network sites.

The dissipation mechanism described by Condition II finds the physics analogy [28] in various transport phenomena, such as diffusion, heat conduction, or electric current, and states that the resistance to social mobility is linearly proportional to the number of identical resisting elements. In particular, the migration flux between sites $i$ and $j$ decreases with the number of intermediate links $R_{ij}$ between them as $J_{ij} \sim 1/R_{ij}$. For a uniform spatial distribution of sites $R_{ij}(r) \sim r$. Combined with the Gauss' mechanism, this results in the migration filed $E_i \sim Q_i^+/r^2$ and the corresponding migration flow $F_{ij} \sim Q_i^+ Q_j^-/r_{ij}^2$, so that the migration exponent $\gamma = 2$.

The combination of the conservation and dissipation mechanisms determines the $1/N(r)$ scaling behaviour of the migration flow. This suggests that the flow from any site is distributed uniformly between all sites within a certain area centred at that site. For the "ideal" network, whose sites are distributed homogeneously in space, this results in the migration flow, which follows the power-2 law scaling, resembling the Newton's gravity law.

**Non-uniform spatial distribution**

Spatial structures of real social networks differ from the ideal one, resulting in deviations from the power-2 law. In particular, for a non-uniform spatial distribution of the network sites, the number of sites located along a perimeter $\ell$ is proportional to $\ell^\lambda$ with $\lambda \neq 1$, which alters the Gauss' mechanism. Furthermore, the amount of sites located along the radius $r$ is proportional to $r^\rho$ with $\rho \neq 1$, which alters the dissipation mechanism. Hence, the migration field $E_i \sim Q_i^+/r^{\lambda+\rho}$, so that $\gamma = \lambda + \rho$, which, in general, differs from 2. This explains the observed variation of the power law migration exponent across different social contexts. In particular, Figure 3 illustrates the variation of $\gamma = \lambda + \rho$ across different urban networks. For residential relocation the variation of $\gamma$ correlates with city polycentricity [3, 29], confirming dependence of the migration scaling behaviour on the spatial structure of the underlying network. In particular, for monocentric cities $\gamma \propto 1.8 - 2.0$, while for non-monocentric cities $\gamma \propto 0.7 - 0.9$, see Table 1.

The spatial distribution of the network sites may, in principle, follow any function (e.g. an exponential). In this case Condition IV should reflect the corresponding network topology, which, in turn, would alter Equation (3). For example, daily home-to-job commute flows take place in a combined suburb-transportation network, with motorway and high-speed public transport serving as "shortcut" links which connect distant suburbs. These shortcuts decrease the "resistance" between distant suburbs, which should be reflected in Condition II. Different network topologies (e.g. scale-free or small-world) are quantified by the corresponding deterrence functions; however, they do not a priori determine the value of the scaling parameter of that function. Hence, the diversity of the actual topological and spatial structures of the underlying social networks explains the variety of the migration scaling behaviour.

**Saturation transitions**

Actual migration flows tend to become independent of the distance at low and high distances, as is evident from Figure 2. These saturation transitions are not captured by any contemporary gravitational model, but can be explained by the proposed mechanism.

The *low-distance* saturation occurs due to a "potential barrier" (Condition IIa), which prevents low-energy agents from migration, limiting the migration flow. This mechanism is analogous to the photoelectric effect in physics and adsorption process in chemistry, where a bound particle requires to overcome a potential barrier in order to become mobile. While existence of such barrier in physics and chemistry is a generic phenomenon, the magnitude (and shape) of this barrier is not universal (i.e. is specific to each particular system), depending on e.g. the structure of the binding site. Similarly, the structure of the potential barrier to social migration may depend on the specific nature of the migration phenomenon. For residential relocation we quantify this barrier by the coordination number of a source site (i.e. the node's degree).

The *high-distance* saturation occurs due to finite size of the network, determining the lower limit for the migration flow. In particular, for a finite network the amount of sited contained within a large enough circle area, stops growing with the radius of that circle, becoming equal to the total amount of sites in the network.

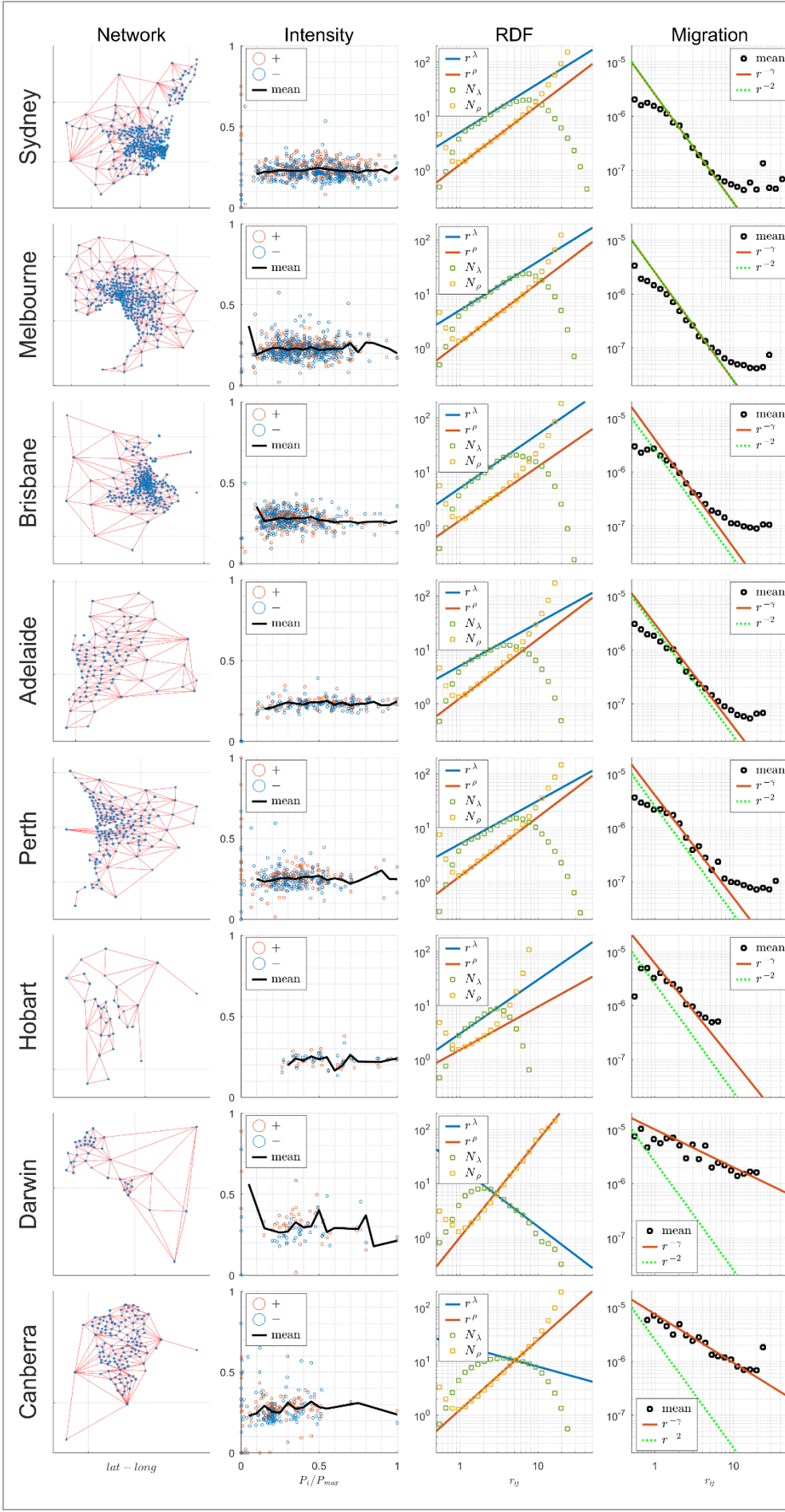

**Figure 3 | Verification of the power-law migration**. Characteristics (columns) for each network, represented by Australian GCAs (rows). *Column 1*: the migration network, with dots representing suburbs and links between them representing their common border. The unit length of the grid corresponds to 0.5 degrees of latitude and longitude. Each GCA has a distinct network structure, along with its size and population (see Table 1). *Column 2*: network's mobility intensity, i.e. the values of $\alpha^+$ and $\alpha^-$ calculated for each suburb (red and blue circles), together with their mean values (black line) as a function of the suburb's population. For each GCA the mean value is approximately independent of the population, justifying Condition III. *Column 3*: the average number of sites as a function of the distance from another site, i.e. the radial distribution function (RDF), for radial ($N_r$, green circles) and perimetric ($N_p$, orange circles) sites, together with the corresponding power law approximation, ($r^\lambda$, blue line) and ($r^\rho$, red line) respectively. The values of the $\lambda$ and $\varrho$ exponents are given in Table 1. The distance is calculated as described in Figure 2. For a significant range of distances each RDF can indeed be approximated by a power law, justifying Condition IV. *Column 4*: the mean of the actual relocation flow quotient $F_{ij}/(P_i^+ P_j^-)$ as a function of the distance (black circles), compared with the predictions of Equation (3) (red line), and with the function $2.5 \cdot 10^{-6}\, r^{-2}$ (green line).

| 8

|          | $P$, mln ppl | $S$, '000 km$^2$ | $N$, nodes | $Q$, '000 ppl | $V$, nodes | $2\theta$, nodes | $\kappa$, x10$^{-6}$ | $\lambda$ | $\rho$ | $\gamma$ |
|---|---|---|---|---|---|---|---|---|---|---|
| Sydney | 4.82 | 12.4 | 312 | 3.6 | 5.12 | 8.37 | 2.5 | 0.9 | 1.1 | 2.0 |
| Melbourne | 4.49 | 10.0 | 309 | 3.3 | 5.50 | 7.60 | 2.5 | 0.9 | 1.1 | 2.0 |
| Brisbane | 2.27 | 15.9 | 236 | 2.6 | 5.18 | 7.80 | 4.0 | 1.0 | 1.0 | 2.0 |
| Adelaide | 1.30 | 3.3 | 110 | 2.8 | 5.25 | 4.68 | 3.0 | 0.8 | 1.1 | 1.9 |
| Perth | 1.94 | 6.4 | 173 | 2.8 | 4.98 | 6.67 | 4.0 | 0.8 | 1.1 | 1.9 |
| Hobart | 0.22 | 1.7 | 35 | 1.5 | 3.94 | 4.42 | 6.0 | 1.0 | 0.8 | 1.9 |
| Darwin | 0.14 | 3.2 | 44 | 0.9 | 4.91 | 3.23 | 10.0 | -1.1 | 1.8 | 0.7 |
| Canberra | 0.40 | 2.4 | 131 | 0.8 | 5.40 | 4.17 | 7.5 | -0.4 | 1.3 | 0.9 |

**Table 1 | Parameters of the migration networks.** The properties (columns) of each network, represented by Australian Greater Capital area (rows). In particular, $P$ – total population, $S$ – total area, $N$ – total number of nodes (suburbs), $Q$ – average source/sink flow, $V$ – average node degree, $2\theta$ – average network path length, $\kappa$ – mobility factor, $\lambda, \varrho$ – exponents of the radial distribution functions, $\gamma$ – migration exponent. $P$, $S$, $N$ are not used in the theory and listed to illustrate the diversity of the underlying networks. $Q$ is calculated from the 2016 Census data of 5-years residential relocation. $V, \theta$ are calculated from the network directly (Figure 3, Column 1). $\lambda, \varrho$ are estimated from the RDF plots (Figure 3, Column 3).

**Summary**

In summary, this paper explains how the distance dependence of social mobility flows emerges from the combination of three key factors: i) the Gauss' theorem; ii) the dissipation mechanism; iii) the spatial distribution of the network sites. The proposed mechanism reveals that the scaling behaviour of social mobility flows is determined by the spatial and topological structure of the underlying social network rather than the explicit geographical distance. In addition to the conventional power-law behaviour, the proposed mechanism also provides a novel explanation for the saturation transitions at low and high distances.

The developed theory is illustrated for human residential relocation in eight Australian Greater Capital areas which differ in population, size, spatial structure and development history. The generic formulation of the theory allows its application to various social contexts with different structures of the underlying social network.

Actual migration flows over the same distance in the same social network may vary across several orders of magnitude. This indicates that there exist individual factors affecting migration flow between particular network sites, such as individual attractiveness [3, 21]. The social gravitational mechanism proposed in this paper does not address individual differences among network sites or migrating agents, describing their mean-field behaviour.

**Acknowledgements**

The author thanks to Mikhail Prokopenko and Bohdan Slavko for the discussion of the preliminary results of the paper. Data are publicly available from the Australian Bureaus of Statistics.

**References**.

[1] Zipf, G. The P1 P2 / D hypothesis: on the intercity movement of persons. *Am. Sociol. Rev*. 11, 677–686 (1946).

[2] Barthélemy, M. Spatial networks. *Physics Reports* 499, 1–101 (2011).

[3] Slavko, B., Glavatskiy, K., Prokopenko, M. City structure shapes directional resettlement flows in Australia. *Scientific Reports* 10, 8235 (2020).

[4] Noulas, A., Scellato, S., Lambiotte, R., Pontil, M. & Mascolo, C. A tale of many cities: universal patterns in human urban mobility. *PLoS ONE* 7, e37027 (2012).

[5] Balcan, D., Colizza, V., Goncalves, B., Hu, H., Ramasco, J.R., Vespignani, A., Multiscale mobility networks and the large spreading of infectious diseases, *Proc. Natl. Acad. Sci.* 106, 21484 (2009).

[6] Jung, W.-S., Wang, F., Stanley, H.E., Gravity model in the Korean highway, *Europhys. Lett.* 81, 48005 (2008).

[7] Viboud, C., Bjornstad, O.N., Smith D.L., Simonsen, L., Miller, M.A., Grenfell, B.T., Synchrony, waves, and spatial hierarchies in the spread of influenza, *Science* 312, 447–451 (2006).

[8] Erlander, S. & Stewart, N. F. The Gravity Model in Transportation Analysis: *Theory and Extensions* (VSP, 1990).

[9] Head, Keith and Mayer, Thierry. "Gravity Equations: Workhorse, Toolkit, Cookbook", *Elsevier's Handbook of International Economics* Vol. 4, 131-195 (2014).

[10] Kaluza, P., Koelzsch, A., Gastner, M.T., Blasius, B., The complex network of global cargo ship movements, *J. R. Soc. Interface* 7, 1093–1103 (2010).

[11] Santos Silva, J.M.C.; Tenreyro, Silvana. "The Log of Gravity". *The Review of Economics and Statistics*. 88 (4): 641–658 (2006).

[12] Deardorff, Alan. "Determinants of Bilateral Trade: Does Gravity Work in a Neoclassical World?" *The Regionalization of the World Economy*. (1998).

[13] Bergstrand, Jeffrey H. "The Gravity Equation in International Trade: Some Microeconomic Foundations and Empirical Evidence". *The Review of Economics and Statistics* 67 (3), 474–481 (1985).

[14] Deville, P. et al. Dynamic population mapping using mobile phone data. *Proc. Natl Acad. Sci*. 111, 15888–15893 (2014).

[15] Krings, G., Calabrese, F., Ratti, C., Blondel, V.D., A gravity model for inter-city telephone communication networks, *J. Stat. Mech.* L07003 (2009).

[16] Lambiotte, R., Blondel, V.D., de Kerchove, C., Huens, E., Prieur, C., Smoreda, Z., Van dooren, P., Geographical dispersal of mobile communication networks, *Physica A* 387, 5317–5325 (2008).

[17] Barbosa, H., Barthelemy, M., Ghoshal, G., James, C. R., Lenormand, M., Louail, T., Menezes, R., Ramasco, J. J., Simini, F., Tomasini, M. Human mobility: Models and applications. *Physics Reports* 734, 1–74 (2018)

[18] Batty, M. The New Science of Cities (MIT Press, 2013).

[19] Bettencourt, L. M. A. The origins of scaling in cities. *Science* 340, 1438–1441 (2013).

[20] Song, C., Koren, T., Wang, P., Barabási, A.-L. Modelling the scaling properties of human mobility. *Nat. Phys*. 6, 818–823 (2010).

[21] Anderson, J.E. Theoretical foundation for the gravity equation, *Amer. Econom. Rev.* 69, 106–116 (1979).

[22] Simini, F., González, M. C., Maritan, A., Barabási, A.-L.. A universal model for mobility and migration patterns. *Nature* 484, 96-100 (2012).

[23] Masucci A. Paolo, Arcaute Elsa, Hatna Erez, Stanilov Kiril and Batty Michael. On the problem of boundaries and scaling for urban street networks *J. R. Soc. Interface*. 1220150763 (2015).

[24] Expert, P., Evans, T. S., Blondel, V. D. & Lambiotte, R. Uncovering space independent communities in spatial networks. *Proc. Natl Acad. Sci.* 108, 7663–7668 (2011).

[25] Roth, C., Kang, S., Batty, M. & Barthélemy, M. Structure of urban movements: Polycentric activity and entangled hierarchical flows. *PLoS ONE* 6, e15923 (2011).

[26] Masucci, A. P., Serras, J., Johansson, A. & Batty, M. Gravity versus radiation models: On the importance of scale and heterogeneity in commuting flows. *Phys. Rev. E* 88, 022812 (2013).

[27] Evans, B. P., Glavatskiy, K., Harré, M., Prokopenko, M. The impact of social influence in Australian real-estate: market forecasting with a spatial agent-based model. *arXiv*:2009.06914 (2021).

[28] Slavko, B., Prokopenko, M., Glavatskiy, K. S. Diffusive resettlement: irreversible urban transitions in closed systems. *Entropy* 23 (1), 66 (2021)

[29] Slavko, B., Glavatskiy, K. S., Prokopenko, M. Revealing configurational attractors in the evolution of modern Australian and US cities. *arXiv*:2011.13597 (2021).